\documentclass[twocolumn,superscriptaddress]{revtex4}
\usepackage{graphicx}
\usepackage{epstopdf}
\usepackage{amsmath}
\usepackage{booktabs}
\usepackage{color}
\usepackage{multirow}
\begin{document}

\title{Non-classical pathways of crystallization in colloidal systems}

\author{John Russo \footnote{e-mail: russoj@iis.u-tokyo.ac.jp}} 
\affiliation{ {Institute of Industrial Science, University of Tokyo, Meguro-ku, Tokyo 153-8505, Japan} }
\affiliation{ {School of Mathematics, University of Bristol, Bristol BS8 1TW, United Kingdom} }
\author{Hajime Tanaka \footnote{e-mail: tanaka@iis.u-tokyo.ac.jp}}
\affiliation{ {Institute of Industrial Science, University of Tokyo, Meguro-ku, Tokyo 153-8505, Japan} }

\begin{abstract}
Colloidal systems offer the ideal conditions to study the nucleation process, both from an experimental viewpoint,
due to their relative large size and long time-scales, and from a modeling point of view, due to the tunability of
their interactions. Here we review some recent works that study the process of colloidal crystallization from a microscopic
perspective. We focus in particular on non-classical pathways to nucleation where the appearance of the solid crystals
involves fluctuations of two (or more) order parameters.
We interpret the non-classical behavior as a decoupling of positional
and orientational symmetry breaking. 
We then consider how the nucleation pathway determines which polymorph is selected upon nucleation from the melt.
Moreover we show how the study of nucleation pathways not only sheds new light on the microscopic mechanism
of nucleation, but also provides important information regarding its avoidance,
thus suggesting a deep link between crystallization and vitrification.
\end{abstract}

\maketitle

\section{Introduction}

Nucleation is an important process both for our understanding of  the dynamics of first-order phase transitions, 
and for its practical applications, ranging from the semiconductor, metal, chemical, 
pharmaceutical, and food industry, and to studies of climate change.
Despite a century of studies on nucleation (for a review see Ref.~\cite{kelton2010nucleation}) the process is still not entirely understood, especially at a microscopic level.
By far the biggest problem with our classical understanding of nucleation is the small size
of the crystalline nuclei involved in the transition (of the order of 10-1000 molecules in usual conditions).
Their small size cast doubts on the use of macroscopic thermodynamic
properties in describing these crystalline nuclei, as for example the use of surface tensions in presence of highly
curved and rough interfaces of small nuclei. At the same time, it makes the observation of crystalline nuclei impractical for most atomic and molecular systems.

A notable class of systems in which the crystallization process is more easily accessible are colloidal suspensions.
The size and the slow dynamics of typical colloidal suspensions, allows
for direct single-particle-level observation with confocal microscopy experiments~\cite{gasser,lowen2007critical,assoud2009crystal,leocmach2012roles,jade_royall}. Moreover, the tunability of their interactions makes
direct comparisons with simple Soft Matter models, like the hard sphere system, possible. Colloidal systems can
thus be considered as an ideal benchmark for testing our understanding of the nucleation theories.
Here we review some of the recent progress made in the field of colloidal crystallization, focusing on non-classical pathways.
These are identified as pathways in which different order parameters take distinct roles during the transition.

\section{Beyond CNT}
The simplest and most general understanding of nucleation is embodied in Classical Nucleation Theory (CNT),
which provides a successful framework to understand and analyze data from nucleation
for a large variety of processes. Yet, its predictions are experimentally very difficult to test, especially given
the exponential nature of the transition, which makes the results very prone to small changes in the experimental
conditions. This difficulty is exemplified in nucleation studies of systems that, despite their simple
interactions, have nucleation rates that are at odds with CNT. The first example is the condensation of Argon~\cite{iland2007argon},
where CNT predictions and experimentally measured rates differ by approximately 26 orders of magnitude.
The second example involves the prototypical colloidal system, hard spheres,
where the discrepancy between predicted nucleation 
rates and experimental measurements is about 10 orders of magnitude.
Numerical simulations found that the nucleation
rate increases dramatically with the colloid volume fraction $\phi$, growing by more than 15 orders of magnitude from
$\phi=0.52$ to $\phi=0.56$, where it has a maximum~\cite{auer2001prediction,zaccarelli,filion,kawasaki,kawasaki2010structural,filion2,schilling_jpcm,valeriani2012compact}. 
On the other hand, experiments found the nucleation rate 
to be much less sensitive to the volume fraction~\cite{iacopini,franke2011heterogeneous},
whose precise determination is also a challenging problem~\cite{royall2013search}.
There have been several attempts at solving this inconsistency, but no consensus has been found yet. Two recent explanations have
emerged. In the first one~\cite{russo2013interplay}, the discrepancy is attributed to experiments, after noting that the gravitational lengths of
many of the colloidal samples are still very short compared to the size of the colloids. The basic idea is that the gravitational field coupled to
long range interactions (such as hydrodynamic interaction), can cause
an increase of density fluctuations in the sample, then triggering crystallization at a higher rate than the gravitation free simulations. The second
explanation instead attributed the discrepancy on simulations, which neglect hydrodynamic interactions~\cite{radu2014solvent}. In the latter simulations,
hydrodynamic interactions have the effect of increasing the nucleation rate of colloids, thus potentially explaining the difference with the experiments.

Classical Nucleation Theory rests on two main assumptions. The first one is the so-called
\emph{capillarity approximation}, which is the assumption that small crystalline nuclei
have the same thermodynamic properties of the bulk solid, like specific volume or surface tension.
Many results contradict these assumption~\cite{oxtoby1998nucleation,granasy1997comparison}.
Several attempts have tried to go beyond the capillarity approximation and an account of these is beyond the scope
of this review; see for example Refs.~\cite{oxtoby1998nucleation,granasy1997comparison,prestipino2012systematic,turci2014solid,lutsko2011communication}.

The second assumption upon which CNT rests is that the process can be described by just one reaction coordinate. 
Under this assumption, all order parameters involved
in the transition proceed simultaneously. Also this assumption has been shown to
fail to capture the dynamics of the transition, especially in simulation works
where different reaction coordinates can be followed during nucleation~\cite{moroni2005interplay,trudu2006freezing,peters2006obtaining,zykova2008irreducible,lechner2011role,russo_hs}.
Also recent theoretical investigations are focusing on the extension of classical nucleation theory to more
than one dimension~\cite{prestipino2014shape,lutsko2015two}. 

Another assumption is that a supercooled liquid is homogeneous and the nucleation happens randomly in space. 
Recent studies showed that this is not the case even for quasi-single-component systems~\cite{tanaka2012bond}.  
Effects of spatial heterogeneity also plays an important role near a critical point if it exists~\cite{ten1997enhancement}. 
Thus, the effects of spatial heterogeneity of the melt should be taken into account.

\section{Two-step nucleation and precursors}
The inability of CNT to distinguish different pathways to crystallization, in which two or more order parameters
evolve independently during the transition, has come to prominence with the discovery of two-step nucleation pathways. 

Two-step crystallization refers to a nucleation pathway where, instead of direct nucleation
from the gas phase (one-step crystallization), the system is found to crystallize inside disordered liquid droplets.
These droplets are usually formed due to the presence of a metastable phase separation~\cite{erdemir2009nucleation,vekilov2010two,gebauer2014pre}.
In colloidal systems, the process has been studied both experimentally~\cite{zhang2007does,savage2009experimental},
and theoretically~\cite{ten1997enhancement,lutsko,soga1999metastable,wedekind2015optimization,haxton2015crystallization}.
The number and variety of solutions that have been found to crystallize from dense liquid precursors lead to the idea that 
two-step crystallization pathways are quite universal crystallization processes. For example, two-step crystallization
processes have been suggested for systems outside the region of stability of the dense fluid phase~\cite{gebauer2014pre,galkin2000control,vekilov2004dense,chattopadhyay2005saxs,sear2009nucleation,lutsko}.

Perhaps the most surprising result is the suggestion that nucleation via dense amorphous regions can occur even in systems where
a dense fluid phase does not exist. The best known example of such a system is given by colloidal models of hard spheres, whose crystallization behaviour has been studied extensively~\cite{pusey,zhu1997crystallization,gasser,martelozzo2002structural,auer2001prediction,bolhuis,filion,filion2}.
For hard spheres, several observations have suggested that the presence of a metastable fluid-fluid demixing transition is not a necessary condition
for a dense precursor-mediated crystallization process~\cite{schope2006two,schope2007preparation,schope2006small,schope2007effect,iacopini,franke2014solidification,schilling,schilling_jpcm}.
The role of density fluctuations, and their priority over structural fluctuations are still subject to investigation,
and for the moment the importance of both contributions cannot be ruled out~\cite{berryman2015early}.
In the following we focus more on works that, for colloidal systems with short range-interactions, point to an alternative nucleation
pathway, where the dominant role is taken by structural fluctuations instead of density fluctuations.

Two fundamental symmetries are broken at the liquid-to-solid transition: translational and orientational symmetry.
Translational order is a measure of the positional order between pair of particles in the system, while orientational
order expressed the degree of angular order between three or more particles in the system.
According to Classical Nucleation Theory,
the crystallization process occurs in one-step, with all the relevant order parameters changing at the same time across the transition.
Two-step crystallization processes, instead,
are at odds with the classical description, as they describe a scenario in which densification precedes structural ordering.
Indeed, also the opposite scenario is possible, with structural order fluctuations preceding the increase of density.

In order to study this process from a microscopic perspective, one should follow the nucleation process from the nucleation event
to the growth of the crystalline nucleus over its critical nucleus size. By using local order parameters, describing the degree of
translational~\cite{russo_hs,mathieu_russo_tanaka} and orientational~\cite{steinhardt,auer2001prediction,auer2004quantitative,lechner,russo_hs} order around individual particles, it is then possible to construct a statistical map of the microscopic
pathway to crystallization. While most of our knowledge comes from molecular simulations on model systems~\cite{kawasaki,kawasaki2010structural,schilling,russo_hs,russo_gcm,russo2013interplay}, recent
confocal microscope experiments have also started to shed light on this process in colloidal systems~\cite{tan2014visualizing,lu2015experimental}.

Several works have suggested that the crystallization in hard sphere-like particles is first triggered by a structural (bond orientational order)
fluctuation, then followed by a densification of the ordered region.
Evidence comes from the study of the correlation length of fluctuations~\cite{kawasaki2007correlation,tanaka,kawasaki,kawasaki2010structural,russo_hs,mathieu_russo_tanaka},
the lifetime of the fluctuations~\cite{kawasaki2007correlation,tanaka,kawasaki2010structural,malins2013identification,malins2013lifetimes,royall2015role},
the radial profile of crystalline nuclei~\cite{oettel2012mode,turci2014solid}, the Landau free energy profile~\cite{russo_hs},
and translational vs orientational curves~\cite{russo_hs,mathieu_russo_tanaka,tan2014visualizing,lu2015experimental}.

Evidence of bond orientational foreshadowing of crystallization has been suggested for colloidal systems both in simulations~\cite{kawasaki,kawasaki2010structural,russo_hs,russo_gcm,mathieu_russo_tanaka,kratzer2015two} and  experiments~\cite{tan2014visualizing,lu2015experimental,mohanty2015multiple}.
Interestingly, the role of structural fluctuations has been investigated also in a variety of different systems, like water~\cite{John_ice0NatMat}, metallic melts~\cite{li2014nucleation,debela2014nucleation}, anisotropic particles~\cite{han2013shape,mahynski2014stabilizing,mondal2015glass},
and polymers~\cite{karayiannis2012spontaneous,hoy2013simple}. The idea of studying crystallization under external fields by studying the
effects of the field on the bond orientational order is also a new direction for future research, whose power has been shown for example
in Ref.~\cite{lander2013crystallization} for the case of crystallization in sheared suspensions.

\section{Polymorph selection}

\begin{figure*}
 \centering
 \includegraphics[width=15cm]{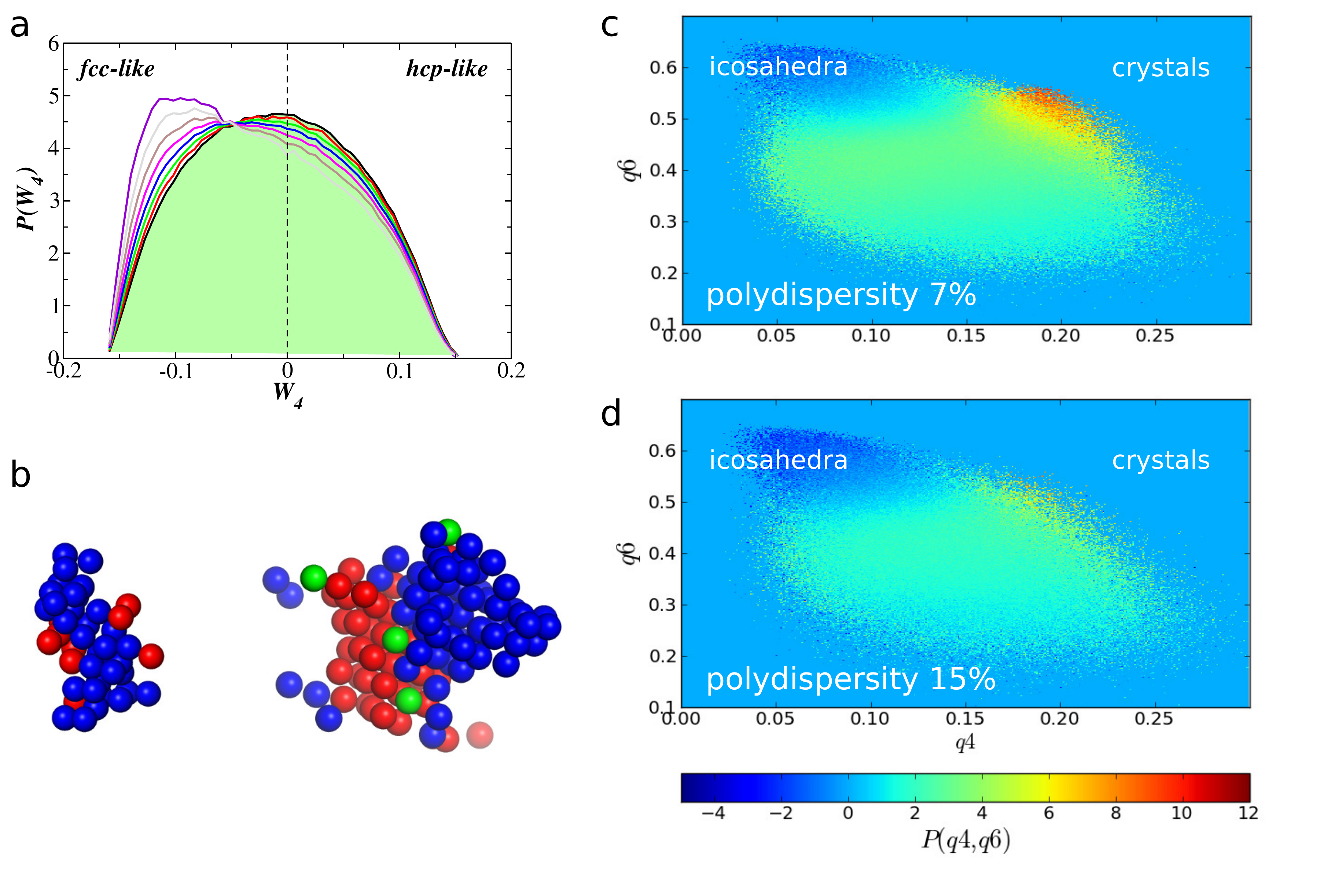}
  \caption{{\bf Polymorphs and glassiness.}
 {\bf a.} Hard spheres at $\beta P\sigma^3=17$: probability distribution of the order parameter $W_4$ for sets of particles
 with increasing values of $Q_6$; the full area is the distribution for the melt, while increasing values of $Q_6$ skew the distribution
 towards the $fcc$ symmetry.
 {\bf b.} Gaussian Core Model at $P=0.05$: two-snapshots of the same nucleus at different times; blue particles are $bcc$ particles, red
 particles are $fcc$ particles.
 {\bf c.} Polydisperse hard spheres at $\beta P\sigma^3=23$: joint probability distribution $P(q_4,q_6)$ for $7\%$ polydispersity.
 {\bf d.} Polydisperse hard spheres at $\beta P\sigma^3=23$: joint probability distribution $P(q_4,q_6)$ for $15\%$ polydispersity.
 Full figures can be found in Refs.~\cite{russo_hs,russo_gcm,mathieu_russo_tanaka}.
 }
 \label{fig:fig1}
\end{figure*}

Polymorph selection plays a fundamental role in many applications, and has been the subject of many investigations~\cite{desgranges2006insights,russo_hs,bolhuis,russo_gcm,mithen2015nucleation}.

The observation that nucleation starts in regions of high bond orientational order 
within the metastable fluid has important consequences in determining which polymorph 
is being nucleated. 
We expect the structure  of a particular crystal to reflect 
the symmetry of the precursor region from which it was nucleated.
This idea was first tested in the hard spheres fluid~\cite{russo_hs}. In hard spheres, the free energy
difference between the fcc and hcp polymorphs is almost indistinguishable.
(around $0.1\%$ of the thermal energy 
in favour of fcc)~\cite{bolhuis_entropy,pronk}. This means that there is a comparable driving force for the nucleation
of either the fcc or hcp crystal. Since the fcc and hcp crystal are polytypes (meaning that they differ only in the
stacking sequence along the direction perpendicular to the hexagonal planes), thermodynamics predicts random stacking 
between fcc and hcp planes (a crystal structure known as rhcp, random close packed structure).
But many studies, both experimental~\cite{gasser,pusey,palberg,jade_royall}
and simulations~\cite{snook,filion,russo_hs}, have found that the fcc phase is much more abundant than the hcp phase.
The preference for fcc nucleation can be understood by looking at the symmetry of the regions where nucleation occurs.
Indeed, a study of the structural properties of supercooled hard spheres, shows that the population of fcc-like precursor regions is bigger than the hcp-like precursor regions~\cite{russo_hs}.
A convenient order parameter which distinguishes between fcc-like and hcp-like orientational symmetry is $W_4$ (for its definition
we refer to Refs.~\cite{leocmach2012roles,russo_hs}): the fcc structure is characterized by negative values of $W_4$, while the hcp structure
by positive values of $W_4$. Fig.~\ref{fig:fig1}a shows the probability distribution of the $W_4$ order parameter for precursor regions 
characterized by increasing values of bond orientational order $Q_6$. While the supercooled melt has a symmetric distribution of
$W_4$, precursor regions progressively peak around negative values of $W_4$, indicating an fcc-like symmetry. The hcp-to-fcc transition would then occur at much longer timescales, due to the negligible
free energy difference between the two different polytypes. 

Polymorph selection was also tested in the Gaussian Core Model (GCM),
a good model for the effective interaction between the centers of mass of polymers dispersed in a good solvent~\cite{likos2006soft}.
The GCM has two different stable crystalline phases, a low pressure fcc-phase and a high-pressure bcc phase.
The nucleation process in the GCM model was first considered in Ref.~\cite{bolhuis}, where it was shown
that the description of the transition is enhanced by taking into account the prestructured particles surrounding the crystalline nucleus.
Ref.~\cite{russo_gcm} then established a clear link between the precursor regions and the crystal phase which was nucleated from them,
also noting a kinetic preference for the bcc phase, even in regions where the stable phase is fcc. Ref.~\cite{mithen2015nucleation}
then found that the nuclei have a mixed nature, not consisting of a single polymorph, and that the kinetic pathway selected during nucleation
persists even when the nucleus is many times above its critical size.
Fig.~\ref{fig:fig1}b shows an example of a mixed-phase nucleus.
All these results suggest that polymorph selection is already made in a supercooled state before 
nucleation starts, and that it is connected with the symmetry of the precursor regions formed by structural fluctuations.

\section{Slow dynamics and external fields}

It is interesting to note that regions of high bond orientational order have not only
been linked to nucleation, but also to its absence, i.e. glass formation \cite{tanaka2012bond}. It was observed that, over the time scale of heterogeneous dynamics, extended regions of high bond orientational order displayed a slower dynamics than disordered regions~\cite{ShintaniNP,leocmach2012roles,kawasaki2007correlation,tanaka,kawasaki2010structural,mathieu_russo_tanaka,russo2015assessing}. Apart from bond-orientational ordered regions, also highly packed regions are 
characterized by slow dynamics~\cite{leocmach2012roles,mathieu_russo_tanaka}. The most notable of these structures is the icosahedral-packing which,
since the pioneering work of Frank, is the archetypal model of amorphous structures.
However, on approaching the glass transition, the icosahedral order does not grow in size, whereas bond-orientational order grows~\cite{leocmach2012roles,mathieu_russo_tanaka}.
A link between slow dynamics and regions of extended bond-orientational order has been found both in polydisperse hard spheres~\cite{tanaka,kawasaki2010structural,leocmach2012roles,mathieu_russo_tanaka}, 
hard disks~\cite{kawasaki2007correlation,kawasaki2011structural,russo2015assessing}, and colloidal ellipsoids~\cite{zheng2014structural}.

In Ref.~\cite{russo_hs} it was shown that, at high density, the population of solid particles is outgrown by particles with icosahedral orders. Even a small fraction of icosahedral particles strongly suppresses the crystallization process~\cite{tanaka2003roles,karayiannis2011fivefold}. The effect becomes more
prominent with the introduction of polydispersity in particle sizes: in Ref.~\cite{mathieu_russo_tanaka} it is shown that the size asymmetry favors
the formation of icosahedral environments, while suppressing the crystalline ones.
Fig.~\ref{fig:fig1}c-d shows the joint probability map of the order parameters $q_4$ and $q_6$. This map is convenient because crystalline
particles are located in the upper-right corner of the map, while icosahedral-like environments in the upper-left corner. We can see that,
going from a system with $7\%$ polydispersity (Fig.~\ref{fig:fig1}c) to a system with $15\%$ polydispersity (Fig.~\ref{fig:fig1}d)
the extent of crystalline regions is highly suppressed, while icosahedral-environments grow.
It is interesting to note that disorder (in this case polydispersity) does not 
always destabilize the crystal. For example, in the case of particles with angular (patchy) interactions, it was shown that angular disorder
does not compromise the crystallizability of the system~\cite{romano2014influence}.
Glasses are thermodynamically unstable, and, despite the slow dynamics, they can crystallize. The study
of both the static~\cite{kawasaki2014structural} and dynamic~\cite{sanz2014avalanches} properties
of the devitrification process holds great promises for the understanding of the glass transition.

Finally, we consider how an external field can couple with the orientational order in the fluid and affect its dynamics and crystallizability.
Ref.~\cite{watanabe_walls} studied this relationship in a vertically vibrated  quasi-two-dimensional granular liquid, and with polydisperse and
bidisperse liquids Brownian simulations. It showed that the walls induce additional glassy order to the fluid, but that this effect
can be encapsulated in a bare correlation length dependent on the distance from the wall. This also suggests a link between bond orientational
order and slow dynamics. Ref.~\cite{russo2013interplay} instead considered the effect of rough walls on colloidal suspensions under the effects
of gravity. It was shown that, while rough walls do not perturb the density field, they strongly suppress bond orientational order up to distances
comparable to the bond-orientational correlation length. This suppression of bond orientational order is then reflected in a strong suppression
of nucleation close to the walls.

Another interesting external field is shear, which has been considered in detail in Ref.~\cite{lander2013crystallization}, showing that
shear suppresses the development of crystallization precursor regions, but favors the growth of crystal nuclei once they are formed. This offers
a compelling explanation of the non-monotonous dependence of the crystallization rate on the shear rate.
The response of colloids to external fields, makes them ideal candidates to uncover their effect on crystallization~\cite{sandomirski2011heterogeneous}

\section{Conclusions}

To conclude, we have reviewed recent progress on non-classical pathways to crystallization in colloidal systems.
Besides advancing our understanding of the nucleation process on a microscopic scale, these studies have
shed light on interesting phenomena, such as polymorph selection, glassiness, and devitrification.
Especially interesting is the study of the effect of external fields on the nucleation process.
Up to now, the case of external walls and uniform shear have been considered, but many other directions are open to investigation.

\end{document}